\documentclass[11pt]{article}

\usepackage[final]{acl}

\usepackage{times}
\usepackage{latexsym}
\usepackage[T1]{fontenc}
\usepackage[utf8]{inputenc}
\usepackage{microtype}
\usepackage{inconsolata}
\usepackage{graphicx}
\usepackage{amsmath}
\usepackage{amssymb}
\usepackage{booktabs}
\usepackage{multirow}

\title{SmallReason-ColBERT: An Ultra-Small Late-Interaction Retriever for Reasoning Intensive Retrieval}

\author{
  \textbf{Abdelrahman Abdallah, Mohammed Ali, Adam Jatowt} \\
  University of Innsbruck, Austria \\
  \texttt{\{abdelrahman.abdallah,adam.jatowt\}@uibk.ac.at}
}

\begin{document}
\maketitle

\begin{abstract}
Reasoning-intensive retrieval remains difficult for small models. Compact public ColBERTs are usually trained on general-purpose corpora and underperform reasoning-tuned 150M+ baselines on BRIGHT~\cite{bright} by several nDCG@10 points. However, no public reasoning-tuned ColBERT exists at edge scale. We introduce \textbf{SmallReason-ColBERT}, a 32M late-interaction retriever that closes much of this gap with three components: a varied-length contrastive warmup on ReasonIR-VL, a hard-negative contrastive polish on merged ReasonIR-HQ and BGE-Reasoner data, and a single-layer per-query-token importance head trained on top of the frozen base. The head is trained with an un-normalised weighted MaxSim score and evaluated with its length-normalised form. In a controlled re-training, replacing this training objective with the symmetric normalised score causes the loss to stall and costs $3.59$ nDCG@10. The full recipe reaches \textbf{21.41} mean nDCG@10 on BRIGHT, within $1.21$ of the 150M Reason-ModernColBERT (22.62) and above all $\le 33$M ColBERTs we evaluate. Through ablations on capacity, initialisation, and score variants, we further show that the learned head outperforms fixed IDF weighting and that simply thresholding the learned gates is harmful.\footnote{\url{https://github.com/DataScienceUIBK/SmallReason-ColBERT}}
\end{abstract}

\section{Introduction}
\label{sec:intro}

Retrieval is now a central component of language-model systems, search engines, and domain-specific question answering pipelines~\cite{croft2010search,abdallah2026rankify,herzig2021open}. As these systems are asked to support longer, more technical, and more reasoning-heavy queries, the retrieval problem increasingly goes beyond lexical overlap or broad semantic similarity: a useful retriever must identify documents that match the intent, constraints, and intermediate concepts of a query~\cite{abdallah2026llm}. This shift has exposed a gap between efficient retrievers that can be deployed widely and larger retrieval models tuned for reasoning-intensive benchmarks.

Late-interaction retrievers~\cite{khattab_colbert_2020,colbertv2} have become a competitive alternative to single-vector dense retrieval. By scoring queries and documents at the granularity of individual tokens and combining them with a per-token MaxSim aggregator, they capture fine-grained interactions that bi-encoders miss while keeping inference costs close to those of bi-encoders. Two recent directions have pushed the family further: small encoders that target on-device retrieval~\cite{mxbai-edge,answeraicolbert}, and reasoning-tuned encoders that target the new generation of reasoning-intensive benchmarks~\cite{reasonmcb,bright}. These directions have not been combined. Public reasoning-tuned ColBERT models are still built at the 150M-parameter scale~\cite{reasonmcb}, while compact public ColBERTs are trained as general-purpose retrievers~\cite{mxbai-edge,answeraicolbert}. Practitioners who need reasoning-aligned retrieval at edge scale therefore face an awkward choice: use a larger reasoning-tuned retriever, or use a compact retriever whose training objective was not designed for reasoning-heavy queries.

In this paper we attempt to close that gap with a three-step recipe applied to a public 32M ColBERT. We first run a varied-length contrastive \emph{warmup} on ReasonIR-VL~\cite{reasonir} to align the model with reasoning-style queries of mixed length. We then \emph{polish} the resulting model with a hard-negative contrastive stage on a merged ReasonIR-HQ and BGE-Reasoner~\cite{bge-reasoner} corpus. The first two steps produce a reasoning-tuned base, but on their own they do not match larger reasoning ColBERTs on BRIGHT. The third and final step adds a single-layer per-query-token \emph{importance head} on top of the frozen base. The head is small but its training is non-trivial: it must be trained against an un-normalised weighted MaxSim and evaluated with the length-normalised score. Replacing this asymmetric training-evaluation pair with the symmetric normalised score collapses the cross-entropy gradient, the head fails to depart from its initialisation, and the controlled BRIGHT mean drops by $3.59$ nDCG@10. The asymmetry is the single most consequential design lever in the recipe.

\paragraph{Contributions.}

\noindent We make the following contributions:
\begin{itemize}
    \item We introduce \textbf{SmallReason-ColBERT}, a 32M late-interaction retriever produced by a three-step recipe: warmup, hard-negative polish, and an importance head. The full recipe reaches \textbf{21.41} mean nDCG@10 on BRIGHT, within $1.21$ of the 150M Reason-ModernColBERT and substantially above all $\le 33$M ColBERTs we evaluate. 
    \item We identify a training objective for the head that uses an un-normalised weighted MaxSim score while retaining length-normalised scores at evaluation time. With this objective the head adds $+1.80$ nDCG@10 over the reasoning-tuned base; with the normalised training score, the head fails to learn useful gates. The recipe is robust to its training hyperparameters across an 18-cell sweep (range $0.42$ nDCG@10). 
    \item  We provide a focused study of score variants on a frozen base. Adding capacity to the head hurts, fixed IDF weighting underperforms the learned gate, and thresholding the learned gates discards useful soft weighting. The single linear head is the best query-weighting variant we tested at this scale. 
    \item  We characterise what the head learns. Contrary to the ``soft-IDF'' interpretation of token importance gates, the learned gate values have near-zero rank correlation with token IDF (Spearman $\rho = -0.02$); the head amplifies tokens that mark query intent rather than tokens that are rare in the corpus.
\end{itemize}

\section{Related Work}
\label{sec:related}

\textbf{Late interaction.} ColBERT~\cite{khattab_colbert_2020} introduced the per-token MaxSim aggregator we build on; ColBERTv2~\cite{colbertv2} added denoised supervision and residual compression that have since become standard. Ja-ColBERTv2.5~\cite{jacolbertv2.5} explored normalisation choices in the distillation loss, including options closely related to but distinct from the train-eval asymmetry we use. \citet{colbertprojections} studied the late-interaction projection head and motivated our investigation of single-vs-multi-layer head choices in \S\ref{sec:abl-headcap}.

\textbf{Compact ColBERT models.} The most directly comparable systems are \texttt{answerai-colbert-small-v1}~\cite{answeraicolbert}, a 33M BERT-MiniLM-based ColBERT, and the \texttt{mxbai-edge-colbert-v0} family~\cite{mxbai-edge} at 17M and 32M parameters built on Ettin~\cite{ettin}. Both are trained on classical IR data and target on-device deployment; neither is reasoning-tuned. Our base re-uses the mxbai-edge backbone but replaces its training distribution and adds the importance head on top.

\textbf{ModernBERT-based late interaction.} GTE-ModernColBERT~\cite{gtemoderncolbert} is the standard 150M ColBERT built on the ModernBERT family~\cite{modernbert}; Reason-ModernColBERT~\cite{reasonmcb} fine-tunes it on ReasonIR-HQ for reasoning-heavy retrieval and is our closest qualitative reference. Our 32M model approaches its BRIGHT performance with a $4.7\times$ smaller backbone and a head of only 129 parameters.

\textbf{Reasoning-heavy retrieval.} BRIGHT~\cite{bright}, RECOR~\cite{ali2026recor} and MM-bright~\cite{abdallah2026mm} introduced the long-form reasoning queries. ReasonIR~\cite{reasonir} contributed the synthetic data we use for warmup and polish, along with the 8B reasoning-tuned single-vector retriever ReasonIR-8B (also a baseline in Table~\ref{tab:main_bright}). BGE-Reasoner~\cite{bge-reasoner} is a complementary reasoning-aligned hard-negative dataset used in our polish stage. A line of work on sparse retrieval, including SPLADE~\cite{formal2021splade,formal2021spladev2}, COIL~\cite{gao2021coil}, and uniCOIL~\cite{lin2021unicoil}, learns token-level scalar weights through MLM-style heads. 

Our head is conceptually related but operates inside a multi-vector late-interaction score rather than a sparse vocabulary projection. Our analysis (\S\ref{sec:analysis}) shows that the learned weights are not soft-IDF, contrary to the typical motivation for SPLADE-family weights. To our knowledge no prior work on late interaction articulates the train-unnormalised / eval-normalised asymmetry we identify in \S\ref{sec:asymmetry}.

\begin{figure*}[t]
\centering
\includegraphics[width=.85\textwidth]{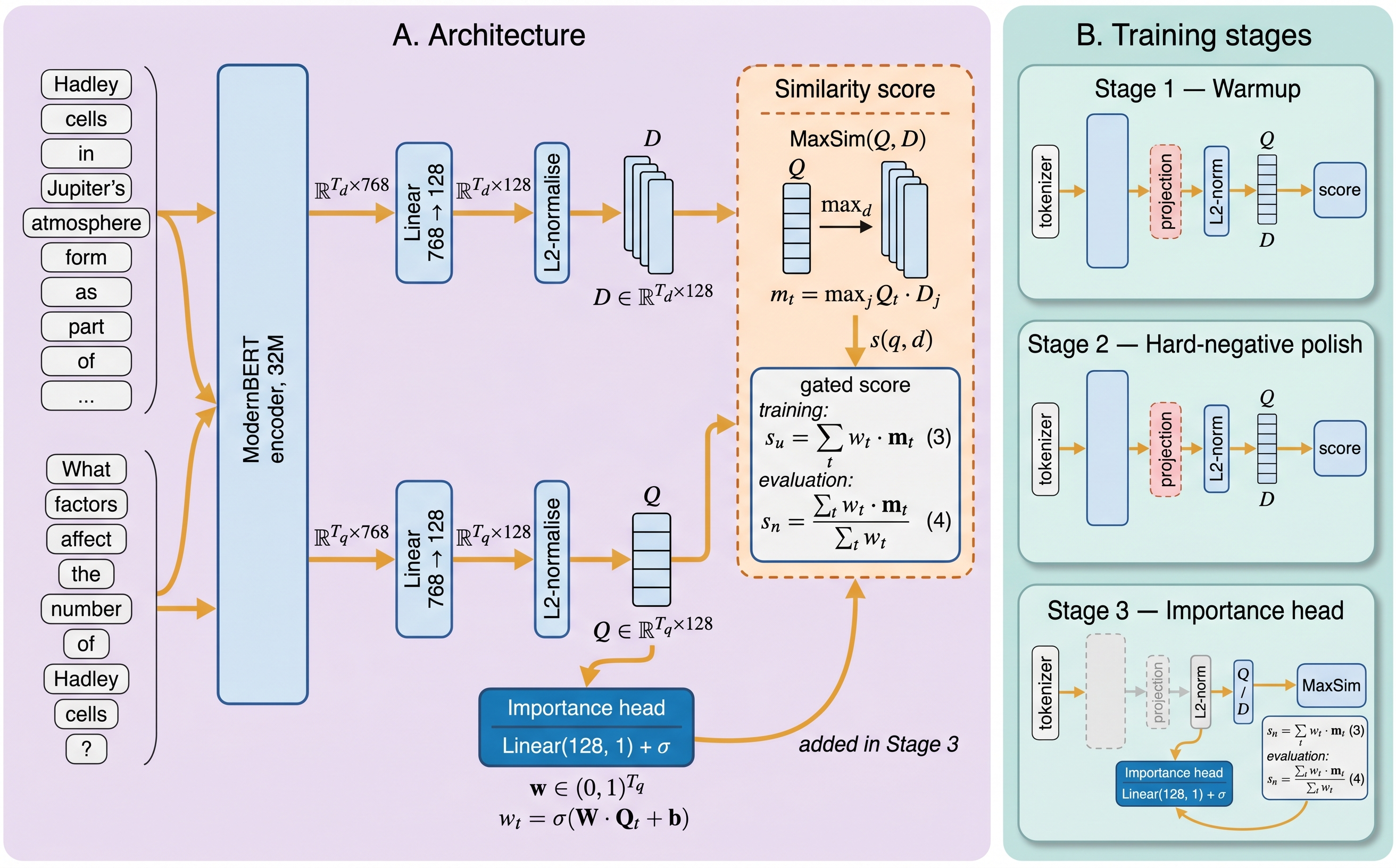}
\caption{Overview of SmallReason-ColBERT. Left: the 32M ModernBERT-based ColBERT encodes query and document tokens, widens the late-interaction projection to 128 dimensions, and applies a query-side importance head before weighted MaxSim scoring. Right: the three-stage recipe first warms up the widened base, then polishes it with hard negatives, and finally trains only the importance head on top of the frozen base.}
\label{fig:overview}
\end{figure*}

% =====================================================================
\section{Method}
\label{sec:method}

Figure~\ref{fig:overview} summarises the architecture and the three training stages. The method has two parts: first we construct a reasoning-tuned 32M ColBERT base by widening and training the projection layer, and then we freeze that base and train a 129-parameter query-side importance head for weighted late interaction.

\subsection{Background: Late Interaction and MaxSim}
\label{sec:background}

A ColBERT~\cite{khattab_colbert_2020,colbertv2} retriever encodes a query $q$ into a sequence of $\ell_2$-normalised vectors $\mathbf{Q} \in \mathbb{R}^{T_q \times d}$ and a document $d$ into $\mathbf{D} \in \mathbb{R}^{T_d \times d}$. Relevance is scored by the MaxSim aggregator:
\begin{equation}
\label{eq:maxsim}
s_\text{MaxSim}(q, d) = \sum_{t=1}^{T_q} \max_{j=1}^{T_d} \mathbf{Q}_t \cdot \mathbf{D}_j.
\end{equation}

Each query token contributes its best match against any document token to the sum. The score is differentiable end-to-end and decomposes additively across query tokens. Standard MaxSim also treats every query token as equally informative, which is the property we revisit in \S\ref{sec:head}.

\subsection{Starting Point and Projection Widening}
\label{sec:start}

We start from \texttt{mxbai-edge-colbert-v0-32m}~\cite{mxbai-edge}, a public 32M ColBERT built on the ModernBERT family~\cite{modernbert,ettin}. Its 3-layer projection emits 64-dimensional token vectors, a deliberately compact choice intended for on-device deployment. On reasoning-heavy splits where many tokens look similar at the surface (variables, operators, theorem names), 64 channels are a tight bottleneck. We therefore widen the final projection \texttt{Linear} from $\text{out\_features} = 64$ to $\text{out\_features} = 128$. The first $64$ rows of the new weight matrix are inherited from the original; the remaining $64$ are initialised from $\mathcal{N}(0, \sigma^2)$ with $\sigma$ set to $10\%$ of the empirical standard deviation of the original weight matrix. The new channels are small enough that the model's MaxSim score is approximately unchanged at step zero, while still being non-zero so that they receive gradient flow during warmup. With a zero initialisation, both query and document vectors would be zero on the new channels, making the dot product and its gradient zero on those dimensions.

\subsection{Reasoning-Tuned Base Training}
\label{sec:warmup}

Going directly from the widened model to hard-negative contrastive training  is unstable: %diverges
the random-initialised new channels inject too much noise for a stable curriculum. We therefore train the base in two stages. First, a varied-length warmup aligns the widened projection with reasoning-style queries of mixed length. Second, a hard-negative distillation stage polishes the model on triples that are closer to the downstream retrieval distribution. The concrete corpora, losses, and training hyperparameters are given in \S\ref{sec:training}; this section only defines the model produced by the stages. After warmup and polish, the result is a 32M reasoning-tuned ColBERT that still uses the plain MaxSim score in Eq.~\ref{eq:maxsim}. We use this model as the frozen base for the importance head introduced next, and refer to it as \emph{the base}. The base's retrieval behaviour is analysed in \S\ref{sec:main}; its role in the method is to provide reasoning-aligned token embeddings while leaving the final per-token aggregation uniform.

\subsection{The Importance Head}
\label{sec:head}

We replace the uniform per-token sum in Eq.~\ref{eq:maxsim} with a learned per-token weighting. A single linear layer scores each query token's $d$-dimensional embedding into a sigmoid-bounded gate:
\begin{equation}
\label{eq:head}
w_t = \sigma(\mathbf{W} \cdot \mathbf{Q}_t + b),\quad \mathbf{W} \in \mathbb{R}^{1 \times d},\ b \in \mathbb{R}.
\end{equation}
For $d = 128$ this head has $129$ trainable parameters, less than $0.001\%$ of the base. The head is the only component trained in this stage; the base is frozen.

\subsection{Train Un-Normalised, Evaluate Length-Normalised}
\label{sec:asymmetry}

The head can be incorporated into the score in two natural ways:
\begin{align}
\label{eq:train-score}
s_\text{u}(q, d) &= \sum_{t=1}^{T_q} w_t \cdot \max_{j=1}^{T_d} \mathbf{Q}_t \cdot \mathbf{D}_j, \\
\label{eq:eval-score}
s_\text{n}(q, d) &= \frac{1}{\sum_t w_t}\sum_{t=1}^{T_q} w_t \cdot \max_{j=1}^{T_d} \mathbf{Q}_t \cdot \mathbf{D}_j.
\end{align}
The un-normalised score $s_\text{u}$ scales linearly with the query length $T_q$; the length-normalised score $s_\text{n}$ has the same scale across queries and lies in $[-1, 1]$.

We train against $s_\text{u}$ but evaluate against $s_\text{n}$. This choice matters for optimisation. If we train against $s_\text{n}$, the per-pair score difference $|\Delta s_\text{n}|$ is bounded by the cosine range of a single token's contribution and the cross-entropy gradient over a positive-negative pair drops to $\sim 5 \times 10^{-3}$ at typical magnitudes. The loss stalls near $\ln 2$ and the gates do not learn useful weights. Replacing the un-normalised training loss with the symmetric normalised one degrades BRIGHT mean nDCG@10 by $3.59$ in the controlled ablation (\S\ref{sec:abl-asym}). Training against $s_\text{u}$ lets gradients scale with query length and gives the head a usable learning signal. Within a single query, the rescaling factor $1 / \sum_t w_t$ is constant across documents, so evaluating against $s_\text{n}$ does not change per-query rankings; it only keeps scores comparable across query lengths.

\subsection{Initialisation: a No-Op at Step Zero}
\label{sec:init}

We initialise $\mathbf{W} = \mathbf{0}$ and $b = 5$, so that at step zero every gate equals $\sigma(5) \approx 0.993$. Under this initialisation $s_\text{n}$ reduces to plain MaxSim up to a per-query rescaling that does not change rankings, so the head implements a no-op against the frozen base before any optimisation. This gives the head a stable starting point rather than forcing it to recover from random token weights. We ablate the choice of $b$ in \S\ref{sec:abl-init}. We refer to the final retriever, the reasoning-tuned base plus the trained importance head, as \textbf{SmallReason-ColBERT}.

\begin{table*}[t]
\centering
\caption{Main BRIGHT results: nDCG@10 ($\times 100$). Best at $\le$33M is \textbf{bold} (ties bolded jointly); best overall on the Mean column is \underline{underlined}. ColBERT family models are re-evaluated under our protocol (uniform query length 256, document length 2048); other baselines are taken from the BRIGHT release and the Reason-ModernColBERT report~\cite{reasonmcb}. The Reason-ModernColBERT entry marked $\dagger$ is the published number with per-split tuned query lengths; under our same-protocol evaluation it scores 21.97.}
\label{tab:main_bright}
\setlength{\tabcolsep}{3.0pt}
\renewcommand{\arraystretch}{1.10}
\resizebox{\textwidth}{!}{%
\begin{tabular}{l c c c c c c c c c c c c c c}
\toprule
\textbf{Model} & \textbf{Params} & \textbf{Mean} & \textbf{Bio} & \textbf{Earth} & \textbf{Econ} & \textbf{Psych} & \textbf{Robo} & \textbf{Stack} & \textbf{Sust} & \textbf{Pony} & \textbf{Leet} & \textbf{AoPS} & \textbf{TheoQ} & \textbf{TheoT} \\
\midrule
\multicolumn{15}{l}{\textit{Sparse / classical IR}} \\
BM25                            & --   & 14.53             & 18.90 & 27.20 & 14.90 & 12.50 & 13.60 & 18.40 & 15.00 &  7.90 & 24.40 &  6.20 & 10.40 &  4.90 \\
\midrule
\multicolumn{15}{l}{\textit{Single-vector dense / instruction-tuned}} \\
BGE                             & 567M & 13.71             & 11.70 & 24.60 & 16.60 & 17.50 & 11.70 & 10.80 & 13.30 &  5.70 & 26.70 &  6.00 & 13.00 &  6.90 \\
E5-Mistral                      & 7B   & 17.93             & 18.60 & 26.00 & 15.50 & 15.80 & 16.30 & 11.20 & 18.10 &  4.90 & 28.70 &  7.10 & 26.10 & 26.80 \\
GritLM                          & 7B   & 20.95             & 24.80 & 32.30 & 18.90 & 19.80 & 17.10 & 13.60 & 17.80 & 22.00 & 29.90 &  8.80 & 25.20 & 21.20 \\
Qwen3-Embedding                 & 7B   & 22.51             & 30.60 & 36.40 & 17.80 & 24.60 & 13.20 & 22.20 & 14.80 &  9.90 & 25.50 & 14.40 & 27.80 & 32.90 \\
ReasonIR-8B                     & 8B   & \underline{24.38} & 26.20 & 31.40 & 23.30 & 30.00 & 18.00 & 23.90 & 20.50 & 10.50 & 35.00 & 14.70 & 31.90 & 27.20 \\
\midrule
\multicolumn{15}{l}{\textit{Multi-vector / late interaction (ColBERT family)}} \\
GTE-ModernColBERT               & 150M & 21.72             & 35.30 & 47.86 & 18.97 & 29.70 & 20.49 & 20.18 & 20.76 & 10.58 & 29.99 &  8.91 & 12.87 &  5.07 \\
Reason-ModernColBERT$^{\dagger}$ & 150M & 22.62             & 33.25 & 41.02 & 24.93 & 30.73 & 21.12 & 20.62 & 20.31 &  8.51 & 31.07 &  9.17 & 19.51 & 11.24 \\
\midrule
\multicolumn{15}{l}{\textit{Compact ColBERT ($\le$33M)}} \\
mxbai-edge-colbert (17M)        & 17M  & 18.60             & 29.98 & 41.98 & 17.76 & 20.29 & 16.65 & 16.87 & 19.54 & 12.32 & 27.72 & 10.22 &  8.09 &  1.82 \\
answerai-colbert-small          & 33M  & 18.49             & 28.92 & 37.99 & 18.64 & 21.09 & \textbf{18.95} & 16.90 & 18.08 & 12.87 & 27.36 &  8.90 &  8.42 & \textbf{3.76} \\
mxbai-edge-colbert (32M)        & 32M  & 18.16             & 28.70 & 42.29 & 17.65 & 21.93 & 18.09 & 16.49 & 18.64 & 12.90 & 16.15 &  9.80 & 12.51 &  2.76 \\

\textbf{SmallReason-ColBERT (no head)} & 32M & 19.61 & 33.16 & 44.28 & \textbf{20.25} & 24.91 & 18.65 & 16.66 & 20.11 & \textbf{22.77} & 17.40 &  4.89 &  9.04 &  3.19 \\

\textbf{SmallReason-ColBERT}    & 32M  & \textbf{21.41} & \textbf{34.17} & \textbf{45.03} & 19.99 & \textbf{24.94} & 18.14 & \textbf{17.21} & \textbf{21.07} & 19.33 & \textbf{29.98} & \textbf{10.29} & \textbf{13.00} & 3.74 \\
\bottomrule
\end{tabular}}
\end{table*}
% =====================================================================
\section{Experimental Setup}
\label{sec:setup}

\subsection{Training Data and Recipe}
\label{sec:training}

\textbf{Warmup data.} The first base-training stage uses ReasonIR-VL~\cite{reasonir}, a varied-length reasoning corpus with approximately 245K triples. The mixture contains both long-context reasoning queries and shorter retrieval pairs, making it a natural curriculum for the widened projection described in \S\ref{sec:start}.

\textbf{Polish data.} The second base-training stage uses a merged hard-negative corpus of ReasonIR-HQ~\cite{reasonir} and BGE-Reasoner-data~\cite{bge-reasoner}, totalling about $2.7$M (query, positive, negative) tuples. Both sources target reasoning-aligned retrieval with cross-encoder-mined hard negatives.

\textbf{Base training.} The warmup stage uses a CachedContrastive loss over in-batch negatives; $8\times$ H100, batch $32$ per GPU, gradient accumulation $4$, learning rate $1\times 10^{-5}$, one epoch, query length $256$, document length $2{,}048$, flash-attention~2~\cite{dao2023flashattention2}, and bf16. Training takes about eight hours. The polish stage uses the same CachedContrastive objective on hard negatives, with the same hardware, batch size, sequence lengths, and precision, a learning rate of $5\times 10^{-6}$, and gradient accumulation $2$. Training takes about $16$ hours.

\textbf{Head training.} The base is frozen and only the 129-parameter importance head is trained. We use AdamW with learning rate $5\times10^{-4}$, batch size $16$, and $3{,}000$ steps on the merged ReasonIR-HQ + BGE-Reasoner triples. Training-time scoring uses $s_\text{u}$ (Eq.~\ref{eq:train-score}); evaluation-time scoring uses $s_\text{n}$ (Eq.~\ref{eq:eval-score}). Head training takes roughly twelve minutes on a single H100.

\subsection{Evaluation}
\label{sec:evaluation}

\textbf{Evaluation benchmark.} We evaluate on \textbf{BRIGHT}~\cite{bright}, a reasoning-heavy retrieval benchmark with 12 splits across coding (LeetCode, Pony), math (AoPS, TheoremQA-questions, TheoremQA-theorems), and StackExchange domains (Biology, Earth Science, Economics, Psychology, Robotics, StackOverflow, Sustainable Living). All numbers are mean nDCG@10 ($\times 100$).

\textbf{Eval protocol.} We use a uniform query length of 256 across all splits except Pony (32) and a document length of 2048 truncated with the model's tokenizer. Scoring uses brute-force MaxSim: for our model, the eval-time length-normalised weighted score $s_\text{n}$ (Eq.~\ref{eq:eval-score}); for the baselines, the plain MaxSim of Eq.~\ref{eq:maxsim}. We re-evaluate every public baseline under the same protocol so that all numbers in Table~\ref{tab:main_bright} are directly comparable; for Reason-ModernColBERT, where the original release used per-split tuned query lengths, we additionally report the published number with the dagger marker.

\textbf{OOD evaluation.} We additionally evaluate on \textbf{OBLIQ-Bench}~\cite{obliq}, which probes oblique queries (stance / intent / irony) where similarity-based retrievers are known to fail. We use this as a deliberate OOD stress-test, not as a target.

\textbf{Baselines.} (i) BM25~\cite{bm25}, with split-level numbers taken from the BRIGHT release. (ii) \texttt{mxbai-edge-colbert-v0-32m}~\cite{mxbai-edge} (32M), the upstream public general-purpose ColBERT and the starting point of our recipe. (iii) \texttt{mxbai-edge-colbert-v0-17m}~\cite{mxbai-edge} (17M). (iv) \texttt{answerai-colbert-small-v1}~\cite{answeraicolbert} (33M, BERT-based; document length capped at 512 to respect its position-embedding limit). (v) \texttt{GTE-ModernColBERT-v1}~\cite{gtemoderncolbert} (150M), the same backbone family as Reason-ModernColBERT but trained on general-purpose data. (vi) \texttt{Reason-ModernColBERT}~\cite{reasonmcb} (150M), built on GTE-ModernColBERT-v1 and fine-tuned on ReasonIR data, our direct quality reference.

\textbf{Hardware.} Base training takes about 24 hours on $8\times$H100 across two nodes; head training takes about twelve minutes on a single H100.

\section{Results}
\label{sec:results}

\subsection{Main BRIGHT Comparison}
\label{sec:main}

Table~\ref{tab:main_bright} reports the main BRIGHT comparison. SmallReason-ColBERT reaches 21.41 mean nDCG@10, $+3.25$ over the upstream 32M base (18.16) and $+1.80$ over the same model without the importance head (19.61). It approaches the 150M reasoning-tuned Reason-ModernColBERT (22.62, published; 21.97 under our same-protocol re-evaluation) and trails the 150M general-purpose GTE-ModernColBERT by 0.31 nDCG@10 on the mean. Three observations follow.

\textbf{The reasoning-tuned base alone improves the upstream on the mean, but remains incomplete.} The base scores 19.61, above the upstream's 18.16. The reasoning-aligned curriculum gains substantially on the StackExchange splits (Biology $+4.46$, Earth $+1.99$, Economics $+2.60$, Psychology $+2.98$) and on Pony ($+9.87$), but still regresses on some short, content-dense coding and math splits where the upstream was already strong (AoPS $9.80\to4.89$, TheoremQA-questions $12.51\to9.04$). The base alone, therefore, is not the headline; it is a precondition for the head to work.
\begin{table}[t]
  \centering
  \caption{\textbf{Classical-IR on NanoBEIR.} Mean nDCG@10 over the NanoBEIR suite; the \textit{No Touche} column drops the
  NanoTouche-2020 split, which is an outlier across all three retrievers.}
  \label{tab:nanobeir}
  \small
  \setlength{\tabcolsep}{6pt}
  \begin{tabular}{l c c}
  \toprule
  \textbf{Model} & \textbf{All} & \textbf{No Touche} \\
  \midrule
  mxbai-edge-32m        & 60.47 & 65.51 \\
  SmallReason, no head  & 60.93 & 65.35 \\
  SmallReason           & 60.00 & 65.00 \\
  \bottomrule
  \end{tabular}
\end{table}

\textbf{The head selectively recovers and extends the reasoning splits.} Adding the head lifts the mean by $+1.80$, dominated by the same coding and math splits that the base had regressed: LeetCode $+12.58$, AoPS $+5.40$, TheoremQA-questions $+3.96$. The full system ends up at or above the upstream on every split; the largest regression relative to the un-headed base is Pony, whose 32-token queries are too short for a per-token gate to discriminate informative from uninformative tokens. Per-token weighting is exactly the mechanism long reasoning queries need, and precisely the mechanism an unweighted MaxSim cannot exploit on a base retrained on a length-skewed mixture.

\begin{table}[t]
\centering
\caption{OBLIQ-Bench out-of-distribution evaluation: Gold nDCG@10 ($\times 100$). We list a representative selection of single-stage retrievers and a lightweight agentic baseline from~\cite{obliq}.}
\label{tab:obliq}
\resizebox{\columnwidth}{!}{%
\setlength{\tabcolsep}{3pt}
\renewcommand{\arraystretch}{1.05}
\begin{tabular}{l c c c c c c}
\toprule
\textbf{Model} & \textbf{Twit.} & \textbf{Wild.} & \textbf{Math} & \textbf{Writ.} & \textbf{Cong.} & \textbf{Mean} \\
\midrule
\textbf{Ours (32M)}     &  0.03 &  0.45 &  8.54 &  6.85 &  2.42 &  3.66 \\
Qwen3-Emb-4B            &  3.20 &  3.10 &  9.50 &  3.30 &  4.00 &  4.62 \\
Qwen3-Emb-0.6B          &  0.80 &  5.90 & 11.60 &  4.60 &  0.60 &  4.70 \\
LateOn (149M)           &  0.40 &  0.20 & 11.20 & 10.50 &  8.30 &  6.12 \\
GPT-5.2 Rewriter        &  6.60 &  6.50 & 14.20 &  1.80 &  8.40 &  7.50 \\
\bottomrule
\end{tabular}}
\end{table}
\textbf{32M with a head approaches 150M ColBERTs.} The gap to Reason-ModernColBERT is $1.21$ nDCG@10 (published) or $0.56$ (same-protocol re-evaluation), and the gap to GTE-ModernColBERT is $0.31$. On the StackExchange splits, our model is within 1 nDCG of Reason-ModernColBERT on most splits. This suggests that reasoning-aligned data combined with a tiny query-side gate can recover much of the gap to a $4.7\times$ larger ColBERT on this class of queries. Substantially larger reasoning-tuned single-vector models (ReasonIR-8B at $24.38$, Qwen3-Embedding at $22.51$) maintain an advantage of one to three nDCG@10 points. Closing that remaining gap may require more capacity or a different inductive bias than a frozen 32M base can provide.

\textbf{Where the gain comes from.} The $+1.80$ mean is dominated by three splits: LeetCode ($+12.58$), AoPS ($+5.40$), and TheoremQA-questions ($+3.96$). All three contain long natural-language statements with a small number of operationally distinguishing tokens, such as function names, math operators, and theorem names. Mild regressions appear on Pony ($-3.44$), whose 32-token queries leave less room for a gate to separate content from context, Robotics ($-0.51$), where queries are already content-dense, and Economics ($-0.26$). The head learns a soft preference over query tokens rather than a hard filter; we characterise this behaviour in \S\ref{sec:analysis}.

\subsection{Out-of-Distribution Evaluation}
\label{sec:ood}

\textbf{NanoBEIR.} A natural concern is that adding a query-side gate trained on reasoning data hurts classical, keyword-driven IR. We evaluate on NanoBEIR~\cite{thakur2021beir} (13 BEIR-derived splits sampled to roughly $50$ queries and 1k to 5k documents per split). Table~\ref{tab:nanobeir} reports the mean nDCG@10 of our model with and without the head, alongside the upstream 32M baseline. The head trades $-0.93$ nDCG@10 on classical IR for the $+1.80$ on BRIGHT. Excluding the unstable Touche-2020 split, where the upstream and our final model both score $0.00$ and the un-headed base scores $7.88$ on $50$ queries, the head is nearly flat: $65.35 \to 65.00$ ($-0.35$). The trade-off is expected. A query-side gate optimised on long reasoning queries does not transfer cleanly to short keyword queries, but the resulting degradation is modest.

\textbf{OBLIQ-Bench.} OBLIQ-Bench~\cite{obliq} probes oblique queries about stance, intent, behavioural failure, writing style, and tip-of-the-tongue retrieval. Similarity-based retrievers are known to struggle in this setting. Our model performs near zero on the descriptive subsets (Twitter $0.03$, WildChat $0.45$) and remains low on tip-of-the-tongue queries (Congress $2.42$). Table~\ref{tab:obliq} reports our model alongside a selection of single-stage retrievers and a lightweight agentic baseline from~\cite{obliq}, all of which outperform our model on the mean despite the absolute scores remaining well below the LLM-level upper bound reported in~\cite{obliq}. This is consistent with OBLIQ's thesis that oblique queries require LLM-level verification rather than embedding similarity. We include it as a deliberate negative result and do not claim that a frozen ColBERT with a per-token gate can replace an LLM reranker for this type of query.

\subsection{Compute}
\label{sec:efficiency}

The head adds little inference cost beyond the ColBERT encoder. For each query token (up to $T_q = 256$) it computes one dot product followed by a sigmoid, which is small compared with the surrounding forward pass. The method still inherits the multi-vector storage and scoring costs of late interaction; our efficiency claim is therefore about parameter count and the additional cost of the head, not about replacing single-vector retrieval. The head trains in roughly twelve minutes on a single H100 with the base frozen, compared with about 24 hours on $8\times$H100 to produce the base itself.
\begin{table}[t]
\centering
\caption{Effect of head initialisation on BRIGHT mean nDCG@10. The default setting is the released model; other rows report controlled deltas from the same default.}
\label{tab:abl_init}
\small
\begin{tabular}{l c c}
\toprule
\textbf{Initialisation} & \textbf{Mean} & \textbf{$\Delta$} \\
\midrule
$b = 5$, $\mathbf{W} = \mathbf{0}$      & \textbf{21.41} & 0.00 \\
$b = 0$, $\mathbf{W} = \mathbf{0}$      & 21.41 & 0.00 \\
$b = 10$                               & 21.30 & $-0.11$ \\
$b, \mathbf{W} \sim \mathcal{N}(0, 1)$ & 21.16 & $-0.25$ \\
\bottomrule
\end{tabular}
\end{table}
\section{Analysis}
\label{sec:ablations}

\subsection{Training Asymmetry}
\label{sec:abl-asym}

The most consequential design choice in the recipe is the training score used for the head. We hold every other choice fixed and replace the un-normalised training loss with the symmetric length-normalised one, so the model trains and evaluates against $s_\text{n}$. The cross-entropy loss stalls near $\ln 2 \approx 0.693$, gates barely depart from their $\sigma(5)\approx 0.99$ initialisation, and BRIGHT mean nDCG@10 drops by $3.59$. Training and evaluating both against $s_\text{u}$ is rank-invariant for per-query metrics like nDCG, since a constant per-query rescaling cannot reorder a single query's documents. We therefore use $s_\text{u}$ for the learning signal and $s_\text{n}$ for length-comparable scores.

\subsection{Initialisation Analysis}
\label{sec:abl-init}

The identity-at-step-zero initialisation ($b = 5$, $\mathbf{W} = \mathbf{0}$) implements a no-op against the un-headed base before training. We compare against three alternatives in Table~\ref{tab:abl_init}. Setting $b = 0$ starts gates at $0.5$ everywhere; the converged result matches $b = 5$ at this seed, suggesting that the near-identity start is useful mainly as a stable initial condition. Setting $b = 10$ saturates the sigmoid and slightly reduces gradient flow through the bias, costing $-0.11$ nDCG. A fully random Gaussian initialisation gives the worst result of the four. The differences are small but consistent.

\subsection{Hyperparameter Robustness}
\label{sec:abl-hp}

To verify that the headline number is not a cherry-picked hyperparameter point, we sweep an $18$-cell grid in $\text{lr} \times \text{steps} \times \text{batch}$, retraining the head in each cell and re-evaluating on full BRIGHT. Across the sweep, all runs stay within a $0.42$ nDCG@10 range and a $0.11$ standard deviation. This supports robustness of the recipe once the un-normalised training loss is in place.

\subsection{Head Capacity Analysis}
\label{sec:abl-headcap}

A natural objection is that a single linear gate is under-capacitated. We replaced the 1-layer head with a 2-layer MLP $\text{Linear}(d, H) \to \text{ReLU} \to \text{Linear}(H, 1)$ for $H \in \{128, 256, 512, 1024\}$. The output layer is zero-initialised with bias $5$, and the hidden layer uses Xavier initialisation so that the step-zero behaviour remains near identity. All four MLPs use the same training schedule and base. Table~\ref{tab:abl_capacity} shows that adding capacity does not improve the result. Every MLP underperforms the 1-layer head, and the largest loses $1.06$ nDCG@10 despite having three orders of magnitude more parameters. The frozen base's $128$-dimensional projection appears to be the bottleneck; a single linear gate already extracts the useful signal, while additional parameters overfit the short fine-tuning.

\begin{table}[t]
\centering
\caption{Effect of head capacity on BRIGHT mean nDCG@10. The 1-layer head is the released model; larger heads report controlled deltas from the same default.}
\label{tab:abl_capacity}
\resizebox{0.45\textwidth}{!}{%
\begin{tabular}{l c c c}
\toprule
\textbf{Head} & \textbf{Params} & \textbf{Mean} & \textbf{$\Delta$} \\
\midrule
1-layer linear           & 129     & \textbf{21.41} & 0.00 \\
2-layer MLP, $H = 128$   & 16{,}641 & 20.68 & $-0.73$ \\
2-layer MLP, $H = 256$   & 33{,}281 & 20.77 & $-0.64$ \\
2-layer MLP, $H = 512$   & 66{,}561 & 20.64 & $-0.77$ \\
2-layer MLP, $H = 1024$  & 133{,}121 & 20.35 & $-1.06$ \\
\bottomrule
\end{tabular}}
\end{table}

\subsection{Alternative Score Formulations}
\label{sec:dead-ends}

We compare the learned head against three closely related score variants on the same frozen base. Table~\ref{tab:score_formulations} summarizes the results. The comparison clarifies the role of the importance head: Top-$K$ pooling is a small orthogonal improvement, fixed IDF is useful but weaker than the learned gate, and hard-thresholding the learned gates loses the soft weighting signal.

\begin{table}[t]
\centering
\caption{Alternative scoring variants on the frozen reasoning-tuned base. All values are BRIGHT mean nDCG@10. The gate-threshold run covers 10 of the 12 splits (earth science and AoPS are missing); restricted to those same 10 splits the main model scores 20.16 and Top-$K$ ($K=4$) scores 20.37. Mathematical definitions are given in Appendix~\ref{app:scoring}.}
\label{tab:score_formulations}
\resizebox{0.43\textwidth}{!}{%
\begin{tabular}{l c c}
\toprule
\textbf{Variant} & \textbf{Splits} & \textbf{Mean} \\
\midrule
SmallReason-ColBERT (main)          & 12/12 & 21.41 \\
Top-$K$ MaxSim ($K=4$)              & 12/12 & 21.66 \\
Fixed IDF gate                      & 12/12 & 20.06 \\
Gate threshold ($w_t<0.95$)         & 10/12 & 16.02 \\
\bottomrule
\end{tabular}}
\end{table}

\textbf{Top-$K$ MaxSim.} Replacing $\max_j \mathbf{Q}_t \cdot \mathbf{D}_j$ with the mean of the top $K$ values per query token~\cite{colbertprojections} reaches $21.66$ at $K = 4$, a $+0.25$ improvement over our recipe at $K = 1$. This modification changes how each query token pools document evidence, while our contribution changes how query tokens are weighted. We therefore treat Top-$K$ pooling as an orthogonal scoring variant rather than the main contribution.

\textbf{Fixed IDF anchoring.} Replacing the learned gate with a fixed weighting $w_t = \mathrm{IDF}(\text{token}_t)$ reaches $20.06$, $+0.45$ over the un-headed base but $-1.35$ below the learned head. The IDF prior is informative on its own; the learned head finds a different and stronger signal, as we analyse in \S\ref{sec:analysis}.

\textbf{Gate thresholding.} Setting gates below $0.95$ to zero at evaluation time reduces performance to $16.02$. The head's contribution lies in relative weighting within a query, not in absolute calibration. Many useful query tokens receive moderate weights, so hard pruning removes signal that the weighted sum still needs.

% =====================================================================
\section{Analysis: What Does the Head Learn?}
\label{sec:analysis}

A natural reading of token-level importance gates is that they approximate inverse document frequency, amplifying rare tokens and suppressing common stopwords. We test this hypothesis directly. We compute the gate value $w_t$ for every query token across all 12 BRIGHT splits ($\sim 200{,}000$ tokens in total), pair each token with its IDF in the BRIGHT/earth-science corpus (a representative full-corpus index), and measure the rank correlation between the two.

Table~\ref{tab:gate_stats} summarises the per-split gate distribution. Across all 12 splits the mean gate sits between $0.43$ and $0.47$ with a standard deviation of about $0.10$, and the distributions are unimodal with a long tail towards $1$. The head is doing soft modulation, rather than hard selection. The gate rarely zeros out a token, but it routinely doubles the contribution of certain tokens relative to others within the same query.
\begin{table}[t]
\centering
\caption{Gate statistics on BRIGHT query tokens (12 splits, $\sim 200$K tokens). Values are computed over $\sigma(\mathbf{W} \mathbf{Q}_t + b)$ for every query token. Distributions are unimodal and consistent across splits.}
\label{tab:gate_stats}
\small
\begin{tabular}{l c c c}
\toprule
\textbf{Statistic} & \textbf{Min} & \textbf{Median} & \textbf{Max} \\
\midrule
Per-split mean      & 0.43 & 0.45 & 0.47 \\
Per-split std       & 0.09 & 0.10 & 0.11 \\
Token-level minimum & 0.31 & --   & --   \\
Token-level maximum & --   & --   & 1.00 \\
\bottomrule
\end{tabular}
\end{table}
Across the matched (gate, IDF) pairs the Spearman rank correlation is $\rho = -0.02$ ($p = 0.026$), and the Pearson correlation is $-0.04$ ($p < 10^{-5}$). Both are statistically detectable on $\sim 12{,}000$ pairs but practically zero. The head is \emph{not} learning a soft IDF schedule. This finding is consistent with the IDF-anchoring ablation (\S\ref{sec:dead-ends}): a fixed IDF weighting reaches $20.06$ ($+0.45$ over the un-headed base), but the learned head reaches $21.41$ ($+1.80$). Two distinct signals coexist; the IDF prior is informative on its own, but the head finds an additional and stronger signal that has near-zero correlation with token rarity. Qualitative examples are provided in Appendix~\ref{app:qualitative}.

\section{Conclusion}
\label{sec:conclusion}

We presented \textbf{SmallReason-ColBERT}, a 32M late-interaction retriever for reasoning-heavy retrieval. The model is trained in three stages: a varied-length contrastive warmup, a hard-negative distillation polish, and a single-layer per-query-token importance head trained with an un-normalised weighted MaxSim objective. It reaches $21.41$ mean nDCG@10 on BRIGHT, within $1.21$ nDCG of the $4.7\times$-larger Reason-ModernColBERT and substantially above all $\le 33$M ColBERTs we evaluate. In the controlled ablation, replacing the asymmetric training objective with the symmetric normalised score costs $3.59$ nDCG@10; the cross-entropy gradient collapses and the head fails to learn. Through analyses of capacity, initialisation, and score variants we show that the minimal head is a design optimum at this scale: adding capacity to the head hurts, fixed IDF is weaker than the learned gate, and hard thresholding removes useful soft-weighting signal. Analysis of the learned gate values shows that the head is not learning soft-IDF; it amplifies tokens that mark query intent rather than tokens that are rare. 

\section*{Acknowledgments}
The authors would like to acknowledge the financial support provided by the Austrian Research Agency (FFG) for the project “AI Enabled Sustainability Jurisdiction Demonstrator” (project No. 915229). The computational results presented in this work have been achieved using the MUSICA cluster, part of the Austrian Scientific Computing (ASC) infrastructure.

\section*{Limitations}
\label{sec:limitations}

\textbf{Scale of the backbone.} The recipe is developed and validated at the 32M scale that this paper targets. We do not claim that the same three-step recipe transfers without modification to substantially smaller backbones, to larger ColBERT models, or to non-ColBERT late-interaction families; the choice of widening, learning rates, and training-data mixtures was tuned to this backbone, and other scales may need different settings.

\textbf{Frozen-base assumption.} The importance head is trained on top of a frozen reasoning-tuned base. We did not explore joint fine-tuning of the base and the head, nor whether the head's contribution persists if the base is further updated. Both are natural follow-ups but lie outside the scope of this paper.

\textbf{Inference cost of late interaction.} Although the head itself adds negligible compute, the model still inherits the multi-vector storage and scoring costs of ColBERT-style late interaction. Our efficiency claim is about parameter count and the marginal cost of the head, not about matching the storage or query-time cost of single-vector retrieval.

\textbf{Reliance on synthetic teacher data.} The warmup and polish stages train on ReasonIR-VL, ReasonIR-HQ, and BGE-Reasoner, whose hard negatives were mined with cross-encoders. Any systematic biases in that negative mining or in the synthetic query distribution can propagate into the resulting retriever, and we have not audited the teacher behaviour beyond using the released data as-is.

\bibliography{custom}

@inproceedings{khattab_colbert_2020,
  title={Colbert: Efficient and effective passage search via contextualized late interaction over bert},
  author={Khattab, Omar and Zaharia, Matei},
  booktitle={Proceedings of the 43rd International ACM SIGIR conference on research and development in Information Retrieval},
  pages={39--48},
  year={2020}
}

@inproceedings{colbertv2,
  title={Colbertv2: Effective and efficient retrieval via lightweight late interaction},
  author={Santhanam, Keshav and Khattab, Omar and Saad-Falcon, Jon and Potts, Christopher and Zaharia, Matei},
  booktitle={Proceedings of the 2022 Conference of the North American Chapter of the Association for Computational Linguistics: Human Language Technologies},
  pages={3715--3734},
  year={2022}
}

@misc{jacolbertv2.5,
  title={Jacolbertv2. 5: Optimising multi-vector retrievers to create state-of-the-art japanese retrievers with constrained resources},
  author={Clavi{\'e}, Benjamin},
  journal={Journal of Natural Language Processing},
  volume={32},
  number={1},
  pages={176--218},
  year={2025},
  publisher={The Association for Natural Language Processing}
}

@misc{colbertprojections,
  title={Simple Projection Variants Improve ColBERT Performance},
  author={Clavi{\'e}, Benjamin and Lee, Sean and Takehi, Rikiya and Shakir, Aamir and Kato, Makoto P},
  journal={arXiv preprint arXiv:2510.12327},
  year={2025}
}

@misc{answeraicolbert,
  title  = {Small but Mighty: Introducing answerai-colbert-small},
  author = {Clavi{\'e}, Benjamin},
  year   = {2024},
  month  = aug,
  url    = {https://www.answer.ai/posts/2024-08-13-small-but-mighty-colbert.html}
}

@misc{mxbai-edge,
  title={Fantastic (small) retrievers and how to train them: mxbai-edge-colbert-v0 tech report},
  author={Takehi, Rikiya and Clavi{\'e}, Benjamin and Lee, Sean and Shakir, Aamir},
  journal={arXiv preprint arXiv:2510.14880},
  year={2025}
}

@misc{gtemoderncolbert,
title={GTE-ModernColBERT},
author={Chaffin, Antoine},
url={https://huggingface.co/lightonai/GTE-ModernColBERT-v1},
year={2025}
}

@misc{reasonmcb,
  title  = {Reason-{ModernColBERT}},
  author = {Chaffin, Antoine},
  year   = {2025},
  url    = {https://huggingface.co/lightonai/Reason-ModernColBERT}
}

@inproceedings{abdallah2026rankify,
  title={Rankify: A comprehensive python toolkit for retrieval, re-ranking, and retrieval-augmented generation},
  author={Abdallah, Abdelrahman and Piryani, Bhawna and Mozafari, Jamshid and Herzinger, Andreas and Holdcroft, Jamie and Jatowt, Adam},
  booktitle={Proceedings of the 64th Annual Meeting of the Association for Computational Linguistics (Volume 3: System Demonstrations)},
  pages={208--219},
  year={2026}
}

@inproceedings{abdallah2026mm,
  title={Mm-bright: A multi-task multimodal benchmark for reasoning-intensive retrieval},
  author={Abdallah, Abdelrahman and Mounis, Mohamed Darwish and Abdalla, Mahmoud and Kasem, Mahmoud SalahEldin and Senussi, Mostafa Farouk and Mahmoud, Mohamed and Ali, Mohammed and Jatowt, Adam and Kang, Hyun Soo},
  booktitle={Proceedings of the 32nd ACM SIGKDD Conference on Knowledge Discovery and Data Mining V. 2},
  pages={8604--8612},
  year={2026}
}

@inproceedings{ali2026recor,
  title={Recor: Reasoning-focused multi-turn conversational retrieval benchmark},
  author={Ali, Mohammed and Abdallah, Abdelrahman and Agarwal, Amit and Patel, Hitesh Laxmichand and Jatowt, Adam},
  booktitle={Findings of the Association for Computational Linguistics: ACL 2026},
  pages={2688--2723},
  year={2026}
}

@misc{modernbert,
   title={Smarter, better, faster, longer: A modern bidirectional encoder for fast, memory efficient, and long context finetuning and inference},
  author={Warner, Benjamin and Chaffin, Antoine and Clavi{\'e}, Benjamin and Weller, Orion and Hallstr{\"o}m, Oskar and Taghadouini, Said and Gallagher, Alexis and Biswas, Raja and Ladhak, Faisal and Aarsen, Tom and others},
  booktitle={Proceedings of the 63rd Annual Meeting of the Association for Computational Linguistics (Volume 1: Long Papers)},
  pages={2526--2547},
  year={2025}
}

@misc{ettin,
   title={Seq vs seq: An open suite of paired encoders and decoders},
  author={Weller, Orion and Ricci, Kathryn and Marone, Marc and Chaffin, Antoine and Lawrie, Dawn and Van Durme, Benjamin},
  journal={arXiv preprint arXiv:2507.11412},
  year={2025}
}

@misc{dao2023flashattention2,
    title={Flashattention-2: Faster attention with better parallelism and work partitioning},
  author={Dao, Tri},
  booktitle={International Conference on Learning Representations},
  volume={2024},
  pages={35549--35562},
  year={2024}
}

@misc{bright,
 title={Bright: A realistic and challenging benchmark for reasoning-intensive retrieval},
  author={Su, Hongjin and Yen, Howard and Xia, Mengzhou and Shi, Weijia and Muennighoff, Niklas and Wang, Han-yu and Haisu, Liu and Shi, Quan and Siegel, Zachary and Tang, Michael and others},
  booktitle={International Conference on Learning Representations},
  volume={2025},
  pages={48941--48991},
  year={2025}
}

@inproceedings{reasonir,
  title={Reasonir: Training retrievers for reasoning tasks},
  author={Shao, Rulin and Qiao, Rui and Kishore, Varsha and Muennighoff, Niklas and Lin, Xi Victoria and Rus, Daniela and Low, Bryan Kian Hsiang and Min, Sewon and Yih, Wen-tau and Koh, Pang Wei and others},
  journal={arXiv preprint arXiv:2504.20595},
  year={2025}
}

@article{bge-reasoner,
   title={Reasonembed: Enhanced text embeddings for reasoning-intensive document retrieval},
  author={Chen, Jianlyu and Lan, Junwei and Li, Chaofan and Lian, Defu and Liu, Zheng},
  journal={arXiv preprint arXiv:2510.08252},
  year={2025}
}

@inproceedings{thakur2021beir,
  title={Beir: A heterogenous benchmark for zero-shot evaluation of information retrieval models},
  author={Thakur, Nandan and Reimers, Nils and R{\"u}ckl{\'e}, Andreas and Srivastava, Abhishek and Gurevych, Iryna},
  journal={arXiv preprint arXiv:2104.08663},
  year={2021}
}

@misc{obliq,
title={OBLIQ-Bench: Exposing Overlooked Bottlenecks in Modern Retrievers with Latent and Implicit Queries},
  author={Tchuindjo, Diane and Shah, Devavrat and Khattab, Omar},
  journal={arXiv preprint arXiv:2605.06235},
  year={2026}
}

@inproceedings{formal2021splade,
  title={Splade: Sparse lexical and expansion model for first stage ranking},
  author={Formal, Thibault and Piwowarski, Benjamin and Clinchant, St{\'e}phane},
  booktitle={Proceedings of the 44th International ACM SIGIR Conference on Research and Development in Information Retrieval},
  pages={2288--2292},
  year={2021}
}

@misc{formal2021spladev2,
 title={SPLADE v2: Sparse lexical and expansion model for information retrieval},
  author={Formal, Thibault and Lassance, Carlos and Piwowarski, Benjamin and Clinchant, St{\'e}phane},
  journal={arXiv preprint arXiv:2109.10086},
  year={2021}
}

@inproceedings{gao2021coil,
  title={COIL: Revisit exact lexical match in information retrieval with contextualized inverted list},
  author={Gao, Luyu and Dai, Zhuyun and Callan, Jamie},
  booktitle={Proceedings of the 2021 Conference of the North American Chapter of the Association for Computational Linguistics: Human Language Technologies},
  pages={3030--3042},
  year={2021}
}

@misc{lin2021unicoil,
  title={A few brief notes on deepimpact, coil, and a conceptual framework for information retrieval techniques},
  author={Lin, Jimmy and Ma, Xueguang},
  journal={arXiv preprint arXiv:2106.14807},
  year={2021}
}

@article{bm25,
  title={The probabilistic relevance framework: BM25 and beyond},
  author={Robertson, Stephen and Zaragoza, Hugo},
  volume={4},
  year={2009},
  publisher={Now Publishers Inc}
}

@article{abdallah2026llm,
   title={Are LLM-Based Retrievers Worth Their Cost? An Empirical Study of Efficiency, Robustness, and Reasoning Overhead},
  author={Abdallah, Abdelrahman and Holdcroft, Jamie and Ali, Mohammed and Jatowt, Adam},
  journal={arXiv preprint arXiv:2604.03676},
  year={2026}
}

@inproceedings{herzig2021open,
  title={Open domain question answering over tables via dense retrieval},
  author={Herzig, Jonathan and M{\"u}ller, Thomas and Krichene, Syrine and Eisenschlos, Julian},
  booktitle={Proceedings of the 2021 Conference of the North American Chapter of the Association for Computational Linguistics: Human Language Technologies},
  pages={512--519},
  year={2021}
}

@book{croft2010search,
  title={Search engines: Information retrieval in practice},
  author={Croft, W Bruce and Metzler, Donald and Strohman, Trevor and others},
  volume={520},
  year={2010},
  publisher={Addison-Wesley Reading}
}

\appendix
\section{Qualitative Gate Examples}
\label{app:qualitative}

Inspecting the per-token gates on individual BRIGHT queries clarifies what the head amplifies. On natural-language earth-science queries, the gate is a narrow soft modulation around its per-split mean of $0.43$--$0.47$, with a small saturated tail. On a query about Hadley cells in Jupiter's atmosphere, for example, the gate places ``to'' and ``What'' at $1.00$, while the content tokens ``Hadley'' ($0.49$), ``Jupiter'' ($0.52$) and ``cells'' ($0.41$--$0.56$) sit near the mean. Saturation is rare and is not exclusively tied to intent markers: across the five sampled queries only three tokens exceed $0.90$, and one of them is the content noun ``map''; sentence-final ``?'' ranges from $0.43$ to $0.68$. The head therefore acts as a soft re-weighting over query tokens rather than a selector, and the tokens it amplifies most are not the rare content words an IDF prior would favour --- consistent with the near-zero gate--IDF correlation in \S\ref{sec:analysis}.

\section{Scoring Functions}
\label{app:scoring}

This appendix spells out the scoring functions used in the main model and in Table~\ref{tab:score_formulations}. Let
\[
M_t(q,d) = \max_{j \in \{1,\ldots,T_d\}} \mathbf{Q}_t \cdot \mathbf{D}_j
\]
be the standard ColBERT MaxSim value for query token $t$ against document $d$.

\paragraph{Plain MaxSim.}
The un-headed base uses the standard additive ColBERT score:
\begin{equation}
s_{\mathrm{plain}}(q,d) = \sum_{t=1}^{T_q} M_t(q,d).
\end{equation}
This is the path used by the plain baseline evaluator: after computing the query-token-by-document-token similarity matrix, the implementation takes a maximum over document tokens and sums over query tokens.

\paragraph{Learned weighted MaxSim.}
SmallReason-ColBERT adds a query-side scalar gate
\begin{equation}
w_t = \sigma(\mathbf{W}\mathbf{Q}_t + b),
\end{equation}
where the one-layer head has $\mathbf{W}\in\mathbb{R}^{1\times 128}$ and $b\in\mathbb{R}$. The base encoder is frozen, so gradients update only $\mathbf{W}$ and $b$.

During head training, the code uses the un-normalised weighted score:
\begin{equation}
s_{\mathrm{train}}(q,d) =
\sum_{t=1}^{T_q} w_t M_t(q,d).
\end{equation}
At evaluation time, the code first normalises the weights within each query,
\begin{equation}
\bar{w}_t = \frac{w_t}{\sum_{r=1}^{T_q} w_r + \epsilon},
\end{equation}
and then scores
\begin{equation}
s_{\mathrm{eval}}(q,d) =
\sum_{t=1}^{T_q} \bar{w}_t M_t(q,d)
=
\frac{\sum_t w_t M_t(q,d)}{\sum_t w_t + \epsilon}.
\end{equation}
The small $\epsilon$ is an implementation guard against division by zero. The normalisation factor is constant across candidate documents for a fixed query, so it does not change per-query rankings, but it keeps scores on a comparable scale across queries.

\paragraph{Top-$K$ MaxSim.}
The Top-$K$ variant replaces the single best document-token match with the average of the top $K$ matches for each query token. Let $\operatorname{TopK}_K(\mathbf{Q}_t\mathbf{D}^{\top})$ denote the set of the $K$ largest token similarities for query token $t$. The per-token score is
\begin{equation}
M^{(K)}_t(q,d) =
\frac{1}{K}\sum_{z\in \operatorname{TopK}_K(\mathbf{Q}_t\mathbf{D}^{\top})} z,
\end{equation}
and the corresponding weighted score is
\begin{equation}
s_{\mathrm{top}K}(q,d) =
\sum_{t=1}^{T_q} \bar{w}_t M^{(K)}_t(q,d).
\end{equation}
Table~\ref{tab:score_formulations} reports the best value we observed, $K=4$. This variant is orthogonal to the learned importance head: it changes how each query token pools evidence from document tokens, not how query tokens are weighted.

\paragraph{Fixed IDF gate.}
The fixed-IDF variant replaces the learned gate with a non-parametric token rarity weight. For a token $x_t$, let
\begin{equation}
\operatorname{IDF}(x_t) = \log \frac{N}{\operatorname{df}(x_t)},
\end{equation}
where $N$ is the number of documents in the reference corpus and $\operatorname{df}(x_t)$ is the number of documents containing the token. The score is
\begin{equation}
s_{\mathrm{IDF}}(q,d) =
\frac{\sum_t \operatorname{IDF}(x_t) M_t(q,d)}
{\sum_t \operatorname{IDF}(x_t) + \epsilon}.
\end{equation}
This tests whether the learned head is merely recovering a lexical rarity prior. It improves over the un-headed base, but remains below the learned head.

\paragraph{Gate thresholding.}
The thresholding variant keeps the learned gate values but turns them into a hard mask:
\begin{equation}
\tilde{w}_t =
\begin{cases}
w_t, & w_t \ge 0.95,\\
0, & w_t < 0.95.
\end{cases}
\end{equation}
The score then uses the same normalised weighted form,
\begin{equation}
s_{\mathrm{thr}}(q,d) =
\frac{\sum_t \tilde{w}_t M_t(q,d)}
{\sum_t \tilde{w}_t + \epsilon}.
\end{equation}
Its regression in Table~\ref{tab:score_formulations} shows that the head should be interpreted as a soft weighting mechanism rather than as a calibrated token selector.

\end{document}